\newif\ifarxiv\arxivtrue%

\ifarxiv%
\pdfoutput=1
\fi

\documentclass[runningheads,envcountsect,orivec]{llncs}

\ifarxiv%
\RequirePackage[paperheight=235mm,paperwidth=155mm,textwidth=12.2cm,textheight=19.3cm,inner=47pt]{geometry}
\makeatletter
\def\@citecolor{blue}%
\def\@urlcolor{blue}%
\def\@linkcolor{blue}%

\def\orcidID#1{\href{http://orcid.org/#1}{\protect\raisebox{-1.25pt}{\protect\includegraphics{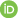}}}}
\makeatother
\fi%

\usepackage{octopus-style}

\begin{document}

\title{Octopus: Practical Equivalence Checking \texorpdfstring{\\}{} of P4 Packet Parsers}

\author{Jort~van~Leenen\corrauthor\orcidlink{0009-0005-9729-4600} \and Tobias~Kappé~\orcidlink{0000-0002-6068-880X}} % chktex 8

\authorrunning{J.~van~Leenen \& T.~Kappé}

\institute{LIACS, Leiden University, The Netherlands\\
\email{j.p.van.leenen@umail.leidenuniv.nl}, \email{t.w.j.kappe@liacs.leidenuniv.nl}%
}

\maketitle

\begin{abstract}
P4 is a domain-specific language for programming protocol-independent packet processors, where packet parsers describe how incoming bit-streams are structured into headers and fields.
Building on work by Doenges et al.\ (2022), we present \textsc{Octopus}, a tool that translates P4 packet parsers into automata and then attempts to (symbolically) check their equivalence.
\textsc{Octopus} produces evidence, either in the form of a bisimulation demonstrating equivalence, or a counterexample bit-stream witnessing a behavioral difference between the two parsers.
In contrast with earlier work, our tool can check equivalence between non-trivial parsers within minutes, on consumer hardware.
We report on the tool’s implementation and evaluate its usability in networking contexts.
\end{abstract}

\section{Introduction}\label{sec:introduction}
\begin{figure}[t]
    \centering
    \includegraphics[
    width=\linewidth,
    alt={%
    Diagram of a UDP packet layout showing a header with four fields, Source Port, Destination Port, Length, and Checksum, followed by a payload that is labeled Data.%
    }
    ]{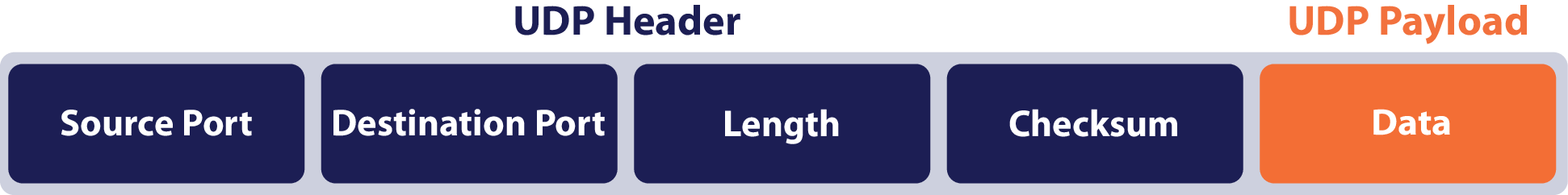}
    \caption{An illustration of the structured information contained in a UDP packet. The left-most four blocks represent the fields comprising the packet's header.
    The right-most block represents the packet's payload.}%
    \label{fig:udp-packet}
\end{figure}

Hosts in a packet-switched network, such as the Internet, communicate by sending packets.
\emph{Switches} (and routers) forward packets to their destinations, indicated in the header field(s).
As an example, consider the illustration of a UDP packet with its header and payload components in \cref{fig:udp-packet}.
A packet switch receives a packet in the form of a stream of bits on an incoming link.
It then has to decide whether the packet is valid, and, if so, where to forward it.
To make this decision, the stream of bits must be parsed into the packet's components and respective fields, which is done by a \emph{packet parser}.

Given the central role of packet parsers, their compliance with functional specifications is critical.
Nonetheless, implementation bugs do occur, and these can lead to incorrect behavior or security vulnerabilities~\cite[Sec.~4.5]{kurose_computer_2022}.
Examples include a set of vulnerabilities called \emph{Ripple20} (\eg, \texttt{CVE-2020-11896}), discovered in the IP stack implementation of many embedded systems, with consequences ranging from remote code execution to denial of service. % chktex 8
This underlines the importance of applying formal methods to ensure parser correctness~\cite{sassaman_security_2013}.

P4~\cite{bosshart_p4_2014} is a domain-specific programming language for \emph{\underline{p}rogramming \underline{p}rotocol-independent \underline{p}acket \underline{p}rocessors}, including the behavior of their packet parsers.
Because P4 is domain-specific, its programs do not have the memory management pitfalls of a general-purpose language like C.
Nevertheless, a P4 program may still contain bugs, either initially, or as a result of modifications aimed at optimizing or refactoring its code.
To address this, we are interested in automatically comparing different versions of the same parser.

\textsc{Leapfrog}~\cite{doenges_leapfrog_2022} is an equivalence checker for P4 packet parsers, implemented in the Rocq proof assistant~\cite{noauthor_rocq_2025}.
It accepts a deeply embedded representation of a subset of P4 parsers, which it attempts to prove equivalent by constructing a symbolic bisimulation~\cite{sangiorgi_introduction_2011} between their automata representations.
Because this latter step is implemented in L\textsubscript{tac}, Rocq's tactic language, it constructs a proof object in the Calculus of Inductive Constructions, which requires a substantial amount of resources.
For example, one particularly large benchmark required more than $400$~GiB of memory and had a runtime of nearly a day.

To alleviate this situation, we present \textsc{Octopus}, a tool that can verify equivalence for the same subset of P4 parsers as \textsc{Leapfrog} --- but being implemented in Python, it does so without the foundational guarantees offered by Rocq.
However, \textsc{Octopus} can run the hardest equivalence query performed by \textsc{Leapfrog} in under a few minutes, on consumer hardware.
To still engender trust in its verdicts, \textsc{Octopus} generates certificates: a bisimulation in the case of equivalence, and a bit-stream witnessing the difference between the parsers otherwise.
Furthermore, \textsc{Octopus} builds its symbolic bisimulation in a forward manner (using strongest postconditions) rather than working backwards (through weakest preconditions), obviating one of \textsc{Leapfrog}'s preprocessing steps~\cite[Sec.~5.1]{doenges_leapfrog_2022}.
Finally, \textsc{Octopus} accepts P4 parsers as input by hooking into \textsc{P4C}~\cite{noauthor_p4langp4c_2025}, the reference compiler of P4, which means that manual translation to \textsc{Leapfrog}'s embedding of parsers is no longer necessary.
These differences entail that \textsc{Octopus} enables \emph{practical} equivalence checking of P4 packet parsers today.

\textsc{Octopus} is available as an open-source project under the MIT license;\footnote{\url{https://github.com/jortvanleenen/Octopus}}
an artifact accompanying this paper is archived permanently on Zenodo.\footnote{\url{https://doi.org/10.5281/zenodo.19832336}}

\section{Parsers in P4}\label{sec:p4-intro}
A packet parser assigns meaning to different parts of its input bit-stream by moving the bits into named \emph{fields}, while building up other metadata during its execution.
These fields are typically organized by protocol into \emph{headers}.
For instance, a UDP packet header (\cref{fig:udp-packet}) can be rendered in P4 as follows.

\begin{lstlisting}[language=P4]
header udp_t { bit<16> src_port;  bit<16> dst_port;
               bit<16> length;    bit<16> checksum; }
\end{lstlisting}

\noindent Specifically, this code defines a header type \texttt{udp\_t}, consisting of four fields named \texttt{src\_port}, \texttt{dst\_port}, \texttt{length}, and \texttt{checksum}, each 16 bits wide.

As packets can be nested, a parser typically defines multiple headers.
Most P4 software switch models then require the programmer to group these separate header containers into a \texttt{struct}, which ultimately holds all headers that the parser might expect to see.
For example, we could combine our UDP header implementation with (simplified versions of) the IP and TCP headers:

\begin{lstlisting}[language=p4, firstnumber=last]
header ip_t      { bit<64> data; }
header tcp_t     { bit<64> data; }
struct headers_t { ip_t ip; udp_t udp; tcp_t tcp; }
\end{lstlisting}

\noindent
A P4 parser is a state machine that moves input data to a header structure and decides whether the packet is accepted for further processing.
It begins in a state called $\start$; two additional states, $\accept$ and $\reject$, represent the eponymous (final) parsing outcomes.
The parser moves between these and user-defined states based on data ingested thus far.
Returning to our example, a basic parser of IP packets containing UDP or TCP payloads may look as follows.

\begin{lstlisting}[language=p4, firstnumber=last,label={lst:p4-example}]
parser Parser(packet_in pkt, out headers_t hdr) {
    state start {
        pkt.extract(hdr.ip);
        transition select(hdr.ip.data[23:20]) {
            0: parse_tcp;
            1: parse_udp;
            _: reject;
        }
    }
    state parse_udp {
        pkt.extract(hdr.udp);
        hdr.udp.length =
            hdr.udp.length[7:0] ++ hdr.udp.length[15:8];
        transition accept;
    }
    state parse_tcp {
        pkt.extract(hdr.tcp);
        transition accept;
    }
}
\end{lstlisting}

\noindent The \texttt{pkt} variable represents the input, and \texttt{hdr} is the variable into which the bit-stream is parsed, also referred to as the \emph{store}.
Each user-defined state starts with an \emph{operation block}, which makes modifications to the store by reading bits.
Roughly speaking, operation blocks contain two kinds of statements:
\begin{enumerate}
    \item
    Calls to \texttt{extract}, which move $n$ bits from the input bit-stream to a field or header of (total) size $n$.
    In the above, this occurs on lines $8$, $16$, and $22$.

    \item
    Assignments, which assign the result of some calculation to a field.
    An example of this occurs on lines $17$--$18$.
    These lines can be interpreted as changing the endianness of the length field, swapping its two bytes.
\end{enumerate}

\noindent
A user-defined state ends with a \emph{transition block}, which decides where to go next.
This can be unconditional (\cf lines 19 and 23), but more commonly the decision depends on the store; for instance, on lines 9--12, the parser decides to advance to \texttt{parse\_tcp}, \texttt{parse\_udp}, or $\reject$ based on bits 20--23 of the \texttt{ip} header. % chktex 1
Here, \texttt{\_} represents a catch-all case.
Besides transition blocks, P4 parsers do not have any primitives for control flow, \ie, there is no \texttt{if}-\texttt{then}-\texttt{else} or \texttt{while}-\texttt{do} construct.

\paragraph{Scope of support}
Like \textsc{Leapfrog}, \textsc{Octopus} is currently limited to parsers where (1)~each state consumes at least one bit of input, (2)~no lookahead is used, and (3)~no header stacks appear.
Assumptions~(1) and~(2) are primarily implementation restrictions and could likely be relaxed with further engineering effort.
However, assumption~(1) is already reasonable in practice, as the P4 specification permits compilers to reject parser cycles that do not advance the cursor~\cite{the_p4_language_consortium_p4_16_2024}.
Assumption~(3) is more fundamental: unbounded header stacks can encode pushdown automata, rendering equivalence undecidable.
Fixed-size header stacks, however, can be encoded using already supported primitives.

\section{Automata Model}\label{sec:automata-model}
Having seen the components that make up a P4 parser, we now outline how they translate to automata.
Abstracting away the nested structure of headers, we represent the defined header fields as a finite set $H$ of \emph{header identifiers}, with each $h \in H$ having a size $|h| \in \mathbb{N}^+$.
When we write $|bv|$ for the length of a bit-vector $bv \in \{0,1\}^*$, we can define a \emph{store} as a member of $St$, given by:
\[St \coloneqq \{ s: H \to \{0,1\}^* \mid \forall h \in H,\, |s(h)| = |h| \}.\]

The states are modeled as a set $Q$, where each $q \in Q$ other than $\accept$ and $\reject$ has an \emph{operation block} $\op(q)$ and a \emph{transition block} $\tz(q)$.
We will refer to the number of bits read by an operation block as its size, denoted by $|\op(q)|$.
As an example, in the P4 parser above, all operation blocks are 64 bits in size.

A \emph{configuration} is a triple $\langle q, s, w \rangle$, which represents a parser in state $q \in Q$, with a store defined by $s \in St$, and the buffer holding the bits $w \in \{ 0, 1\}^*$, with $|w| < |\op(q)|$.
This makes $C$, the set of all configurations, finite.

The \textit{transition function} maps a configuration $\langle q, s, w \rangle$ and a bit $b$ to a next configuration, by either growing the buffer when $|wb| < |\op(q)|$, or executing the operation block $\op(q)$ and transitioning according to $\tz(q)$, clearing the buffer:
\begin{equation*}
  \delta\bigl(\langle q, s, w\rangle, b\bigr) =
  \begin{cases}
    \langle \reject, s, \epsilon \rangle & q \in \{ \accept, \reject \}, \\
    \langle q, s, wb\rangle & |wb| < |\op(q)|, \\
    \langle \sem{\tz(q)}(s^\prime),\; s',\; \epsilon\rangle & |wb| = |\op(q)|,\, s' = \sem{\op(q)}(s,wb).
  \end{cases}
\end{equation*}

In the above, $\sem{\op(q)}$ represents the semantics of the operation block attached to $q$ as a function that takes a store $s$ and a buffer $wb$ of size $|\op(q)|$, and returns an updated store $s'$.
By the same token, $\sem{\tz(q)}$ is a function that determines the next state, which it does based on the newly updated store $s^\prime$.

Putting these parts together, we can represent a P4 parser as a deterministic finite automaton $\langle C, \delta, F \rangle$, which operates on configurations from $C$, has transition function $\delta$, and accepting configurations $F = \{ \langle \accept, s, \epsilon \rangle \mid s \in St \}$.
We do not select an initial configuration, as the semantics of P4 stipulate that store values need not be initialized~\cite{the_p4_language_consortium_p4_16_2024}.
Accepted packets are thus bit-streams accepted by any state of the form $\langle \start, s, \epsilon \rangle$ in the traditional sense.

\section{Equivalence Checking}\label{sec:equiv-check}
We now present the symbolic equivalence-checking algorithm implemented in \textsc{Octopus}, which attempts to decide whether two P4 packet parsers are behaviorally equivalent by constructing a symbolic bisimulation over their automata representations.
Recall that two states in a deterministic automaton have the same language if and only if they are related by a \emph{bisimulation}.

\begin{definition}[Bisimulation]\label{def:bisimulation}
Let $\langle C_i, \delta_i, F_i \rangle$ be an automaton, with $i \in \{1,2\}$.
A relation $R \subseteq C_1 \times C_2$ is a \emph{bisimulation} if, for all $c_1 \mathrel{R} c_2$, we have $c_1 \in F_1$ iff $c_2 \in F_2$, and $\delta_1(c_1, b) \mathrel{R} \delta_2(c_2, b)$ for $b \in \{0,1\}$.
\end{definition}

Thus, equivalence checking reduces to constructing a bisimulation.
Naively, one can do this by starting from an initial pair $(c_1,c_2)$, and repeatedly adding pairs reachable by point-wise transitions.
If we add a pair $(c_1', c_2')$ where $c_1' \in F_1$ but $c_2' \not\in F_2$ (or vice versa), a bisimulation cannot exist; if the process terminates without such a violation, a bisimulation has been constructed.
However, automata constructed from P4 parsers are prohibitively large: even the toy parser from the previous section has roughly $3 \times 2^{63} \times 2^{3 \times 64}$ possible configurations. %, a large part of which is reachable from $\langle \start, s, \epsilon \rangle$ for some $s \in St$.

The key insight to a workable algorithm is that we can represent large sets of concrete pairs of configurations symbolically.
While this methodology is well known, the specific symbolic representation we use stems from \textsc{Leapfrog}~\cite{doenges_leapfrog_2022}.

\begin{definition}[Template-guarded Formula (TGF)]
A \emph{template-guarded formula} (TGF) is a 5-tuple $\gamma = \langle q^<_\gamma, q^>_\gamma, n^<_\gamma, n^>_\gamma, \varphi_\gamma \rangle$, where $q^<_\gamma$ (resp. $q^>_\gamma$) indicates the state of the left (resp.\ right) parser, and $n^<_\gamma$ (resp.\ $n^>_\gamma$) is the number of bits in the left (resp.\ right) parser's buffer.
We refer to the first four components together as a \emph{template}, denoted by $\tau_\gamma = \langle q^<_\gamma, q^>_\gamma, n^<_\gamma, n^>_\gamma \rangle$.

The final component, $\varphi_\gamma$, is a first-order formula over variables $\text{buf}^{\,<}$, representing the buffer (of size $n^<$) in the left configuration, and $\text{st}^<$, representing the store of the left configuration (resp.\ $\text{buf}^{\,>}$ and $\text{st}^>$ in the right configuration).
\end{definition}

Informally, a TGF $\langle q^<_\gamma, q^>_\gamma, n^<_\gamma, n^>_\gamma, \varphi_\gamma \rangle$ denotes pairs of configurations where (1)~the configuration on the left has $q^<_\gamma$ as its state and a buffer with $n^<_\gamma$ bits (and similarly on the right), and (2)~those buffers and stores together satisfy $\phi_\gamma$.

\paragraph{Bisimulation checking}
\cref{algo:bisim} builds a symbolic bisimulation, starting with the TGF $\langle \start, \start, 0, 0, \top \rangle$, which represents all possible pairs of initial configurations for both parsers.
Here, the trivial formula $\top$ does not restrict header fields because P4 defines header fields to be uninitialized~\cite{the_p4_language_consortium_p4_16_2024}.

\begin{algorithm}[bt]
\DontPrintSemicolon%
$\mathcal{W}\gets\{\langle \start, \start, 0, 0, \top\rangle\}$\;
$\mathcal{K}\gets\emptyset$

\While{$\mathcal{W}\neq\emptyset$}{
    $\gamma=\langle q^<_\gamma,q^>_\gamma,n^<_\gamma,n^>_\gamma,\varphi_\gamma\rangle \gets \mathcal{W}.pop()$\;

    \If{$\varphi_\gamma \models \bigvee \{ \phi_{\gamma'} \mid \gamma' \in \mathcal{K} \cup \mathcal{W},\, \tau_{\gamma'} = \tau_\gamma \}$}{\label{algo:checknew}\textbf{continue} \tcp*{no new information}}

    \If{\label{algo:coherence}$\textnormal{\textbf{not}}\ (q^<_\gamma=\accept \iff q^>_\gamma=\accept)$}{\Return counterexample}

    \label{algo:leap}$\ell\gets \min(|\op(q^<_\gamma)| - n^<_\gamma ,\ |\op(q^>_\gamma)| - n^>_\gamma)$ \tcp*{read until buffer full}
    \label{algo:extendbuf}$\varphi'\gets \mathsf{Extend}(\varphi_\gamma,\ell)$ \tcp*{read $\ell$ bits, extend buffers}

    \If{$n^<_\gamma+\ell=|\op(q^<_\gamma)|$}
    {
        $\varphi'\gets \SP^<(\op(q^<_\gamma),\varphi')$;\ $n^<\gets 0$;\ $L\gets\ST^<(\tz(q^<_\gamma))$  \tcp*{left step}
    }
    \lElse{
        $n^<\gets n^<_\gamma+\ell$;\ $L\gets\{(\top, q_\gamma^<)\}$
    }

    \If{$n^>_\gamma+\ell=|\op(q^>_\gamma)|$}
    {
        $\varphi'\gets \SP^>(\op(q^>_\gamma),\varphi')$;\ $n^>\gets 0$;\ $R\gets\ST^>(\tz(q^>_\gamma))$ \tcp*{right step}
    }
    \lElse{
        $n^>\gets n^>_\gamma+\ell$;\ $R\gets\{(\top, q_\gamma^>)\}$
    }

    \ForEach{\text{$(c^<,q'^<)\in L$ \textnormal{\textbf{and}} $(c^>,q'^>)\in R$}}{
        $\mathcal{W}\gets \mathcal{W}\cup\{\langle q'^<,q'^>,n^<,n^>,\varphi' \wedge c^< \wedge c^>\rangle\}$\;
    }

    $\mathcal{K}\gets \mathcal{K}\cup\{\gamma\}$\;
}

\Return $\mathcal{K}$\;
\BlankLine

\caption{Symbolic bisimulation checking over TGFs.}
\label{algo:bisim}
\end{algorithm}

The algorithm then proceeds like the naive approach.
To describe pairs of symbolic configurations, we iteratively compute successor TGFs and disjunctively add those to our candidate symbolic bisimulation.
This process continues until no new TGFs arise (Line~\ref{algo:checknew}, implemented using an SMT solver), in which case a symbolic bisimulation has been obtained, and the parsers are deemed equivalent.
Otherwise, if a TGF is reached where exactly one state in the template is accepting (Line~\ref{algo:coherence}), then we conclude that the parsers are not equivalent.

Calculating successor TGFs is a two-phase process.
First, if a parser lacks sufficient buffered bits to execute the next operation block, the symbolic step consumes input.
It extends the buffer by introducing a fresh buffer variable constrained to be the concatenation of the previous buffer and the newly read bits (\textsf{Extend}).
Second, if enough bits are available, the operation block is executed symbolically; its effect on the formula is captured by taking the strongest postcondition, updating the symbolic relationships between both parsers’ stores and buffers.
\Cref{fig:sp} contains a more formal description of this process.

\begin{figure}[btp]
    \centering
    $\SP^\lessgtr(\varphi, \text{\textbf{prog}})$
    \begin{align*}
        \SP^\lessgtr(\varphi, \text{extract}(hdr))
        &\coloneqq \exists y \in \{0,1\}^{|hdr|} \quad
            \varphi\!\left[
              \frac{y}{\text{st}^\lessgtr.hdr},
              \frac{\text{st}^\lessgtr.hdr \plusplus\, \text{buf}^{\,\lessgtr}}{\text{buf}^{\,\lessgtr}}
            \right],\\
        \SP^\lessgtr(\varphi, \text{hdr.}x = e)
        &\coloneqq \exists y \in \{0,1\}^{|x|} \quad
            \varphi\!\left[\frac{y}{\text{st}^\lessgtr.hdr.x}\right]
            \land \text{st}^\lessgtr . hdr.x = e\!\left[\frac{y}{\text{st}^\lessgtr.hdr.x}\right],\\
        \SP^\lessgtr(\varphi,\, \text{prog}_1;\, \text{prog}_2)
        &\coloneqq \SP^\lessgtr\bigl(\SP^\lessgtr(\varphi,\, \text{prog}_1),\, \text{prog}_2\bigr).
    \end{align*}
    \caption{%
        The strongest postcondition ($\SP^\lessgtr$) of an operation block, with the symbol ${\lessgtr} \in \{<, >\}$ depending on which parser the operation block belongs to.
    }%
    \label{fig:sp}
\end{figure}

After computing the strongest postcondition ($\SP^\lessgtr$), we must symbolically determine the next state.
Updating the template to record this new state is straightforward, but it is equally necessary to record the constraints under which this branch was feasible into the formula.
To correctly encode the semantics of the case structure, we conjoin to the relational formula not only the condition for the selected case, but also the negation of the conditions of all earlier cases that were not taken.
The subroutine to compute these constraints and next states given a transition block is denoted $\ST^\lessgtr$, where, as in $\SP^\lessgtr$, $\lessgtr$ is $<$ or $>$ depending on where the transition block belongs.
If no transition is executed, the symbolic transition leaves the state unchanged under the trivial constraint.

On the whole, the algorithm must terminate, simply because each iteration of the loop either grows the number of configurations (of which there are finitely many) modelled by the TGFs in $\mathcal{K}$, or keeps this quantity constant while shrinking $\mathcal{W}$.
If the algorithm determines that the two given parsers are not equivalent, it will use the SMT solver to generate a model for the last TGF\@.
The existentially quantified buffer variables contained in this model can then be used to extract a counterexample in the form of a packet accepted by one parser, but rejected by the other; we omit the details of this procedure for brevity.
Otherwise, when the main loop terminates, the disjunction of the TGFs in $\mathcal{K}$ represents a symbolic bisimulation on the underlying automaton.

As an optimization, our algorithm adopts the ``leaps'' technique from~\cite{doenges_leapfrog_2022} to improve upon bit-by-bit buffer building by allowing both parsers to advance in one large step until either executes an operation (and thus transition) block.
In~\cref{algo:bisim}, this is implemented by calculating the number of steps to advance on Line~\ref{algo:leap}, before extending the buffers on Line~\ref{algo:extendbuf}.
For a correctness proof of the leaps optimization, please consult~\cite[Lemma~5.6]{doenges_leapfrog_2022-arxiv}.

\paragraph{Certificate validation}
Once a certificate (a set of TGFs) is built, it can be validated by checking whether three conditions hold: (1)~it contains the initial TGF $\langle \start, \start, 0, 0, \top\rangle$; (2)~the states in every TGF agree on acceptance; and (3)~the pure formula in each successor TGF is covered --- where by ``covered'' we mean ``implies the disjunction of TGFs in the certificate with the same template''.

In principle, these checks could be implemented by an external checker, separate from our bisimulation-construction algorithm; in practice, these routines also comprise different parts of our codebase.
Conceptually, however, certificate validation is closely related to certificate building, to the point where the same algorithm applies; indeed, \cref{algo:bisim} can check whether a given set of TGFs constitutes a symbolic bisimulation, simply by initiating the work queue $\mathcal{W}$ to that set --- in this case, running the algorithm will effectively check conditions~(2) and~(3) from above.
This is no surprise, as the main loop of \Cref{algo:bisim} actually checks whether $\mathcal{W}$ is \emph{contained} in a symbolic bisimulation, so instantiating $\mathcal{W}$ to $\{\langle \start, \start, 0, 0, \top\rangle\}$ checks whether a bisimulation satisfying (1) exists.

\section{Evaluation}\label{sec:eval}
\textsc{Octopus} is implemented as a command-line tool in Python.
It accepts either two P4 programs, in which case \textsc{P4C}~\cite{noauthor_p4langp4c_2025} is invoked internally to generate their compiler intermediate representation (IR) JSON, or two IR JSON inputs directly.
This IR is then parsed and interpreted via the automata model described in~\cref{sec:automata-model}, on which~\cref{algo:bisim} is executed.
If successful, \textsc{Octopus} returns a certificate of equivalence in the form of a symbolic bisimulation, represented as a disjunction of TGFs.
Otherwise, it returns a counterexample bit-stream accepted by only one parser, as well as the paths taken by both parsers on this input.
The counterexample is derived from the TGFs at the point of failure, and enables mechanical reproduction, while the paths support manual debugging.

By default, \textsc{Octopus} uses \textsc{Z3}~\cite{hutchison_z3_2008} as its SMT solver.
Profiling shows that solver performance is dominant, with about $80$--$85\%$ of execution time spent in SMT calls.
Although \textsc{Octopus} supports solver portfolios via \textsc{PySMT}~\cite{gario_pysmt_2015}, we observed no performance benefit in our experiments; instead, portfolio management introduced overhead that degraded runtime.
We also evaluated using \textsc{cvc5}~\cite{fisman_cvc5_2022} as solver: on public benchmarks (435 small parser comparisons), performance is similar, likely due to small queries and non-SMT overhead, whereas on larger benchmarks (\eg, \emph{variable-length format~3}) differences become more pronounced, and \textsc{cvc5} can be up to $50\%$ faster.
We nevertheless use \textsc{Z3}, as \textsc{cvc5} triggers a crash in the \textsc{PySMT} library for some benchmarks.

\paragraph{Comparison with \textsc{Leapfrog}.}
We compare \textsc{Octopus} against \textsc{Leapfrog}, first on its own benchmark suite~\cite{doenges_leapfrog_2022}, and then in terms of their trusted base.

\textsc{Leapfrog}'s benchmarks can be divided into two broad categories.
The first consists of representative case studies that capture common parser transformations and design idioms.
The second comprises larger parsers drawn from realistic networking scenarios, as well as a translation-validation benchmark.

We first reproduced \textsc{Leapfrog}'s benchmarks; next, we translated the deeply embedded Rocq representation of the parsers to P4 and ran the same comparisons using \textsc{Octopus}.
All experiments ran on an Intel Xeon E5-2630v3 with $1.4$~TiB of memory to accommodate \textsc{Leapfrog}'s memory footprint, and to ensure comparability. % chktex 8
The results for \textsc{Octopus} are averaged over three runs.

\begin{table}[hbtp]
    \centering
    \caption{%
        Resource usage for each case study in \textsc{Leapfrog}'s benchmark suite under both \textsc{Leapfrog} and \textsc{Octopus}.
        Runtime (Run.) refers to the wall-clock time (in minutes) required for equivalence checking.
        For \textsc{Octopus}, runtime is divided into certificate generation (Gen.) and validation (Val.), both in minutes.
        Memory (Mem.) indicates the peak resident memory, in GiB, over the course of the experiment.
    }
    \setlength{\tabcolsep}{8pt}
    \begin{tabular}{lrrrrr}
    \toprule
    & \multicolumn{2}{c}{\textsc{\textbf{Leapfrog}}}
    & \multicolumn{3}{c}{\textsc{\textbf{Octopus}}} \\
    \textbf{Name} & \shortstack{\textbf{Run.}\\[-1pt]\textbf{(m)}} & \shortstack{\textbf{Mem.}\\[-1pt]\textbf{(GiB)}} &
    \shortstack{\textbf{Gen.}\\[-1pt]\textbf{(m)}} & \shortstack{\textbf{Val.}\\[-1pt]\textbf{(m)}} & \shortstack{\textbf{Mem.}\\[-1pt]\textbf{(GiB)}} \\
    \midrule
     State rearrangement      & $1.13$    & $1.07$   & $0.002$  & $0.002$  & $0.047$ \\
     Variable-length format 2 & $3825.70$ & $391.91$ & $0.268$  & $0.340$  & $0.056$ \\
     Variable-length format 3 & --        & --       & $12.149$ & $10.885$ & $0.252$ \\ % chktex 8
     Header initialization    & $21.56$   & $13.48$  & $0.004$  & $0.003$  & $0.048$ \\
     Speculative extraction   & $5.45$    & $3.26$   & $0.003$  & $0.003$  & $0.047$ \\
     Relational verification  & $1.59$    & $1.65$   & $0.003$  & $0.003$  & $0.048$ \\
     External filtering       & $2.37$    & $2.03$   & $0.003$  & $0.002$  & $0.047$ \\
    \midrule
     Edge                     & $730.12$  & $243.35$ & $0.075$  & $0.081$  & $0.049$ \\
     Service provider         & $7167.03$ & $823.70$ & $0.040$  & $0.041$  & $0.049$ \\
     Datacenter               & $1975.92$ & $382.17$ & $1.176$  & $1.949$  & $0.062$ \\
     Enterprise               & $817.12$  & $66.35$  & $0.887$  & $1.272$  & $0.058$ \\
     Translation validation   & $1035.38$ & $349.15$ & $0.120$  & $0.131$  & $0.055$ \\
    \bottomrule
    \end{tabular}%
    \label{tab:experiment-findings}
\end{table}

\Cref{tab:experiment-findings} summarizes the results.
We implemented certificate validation to run after generation, allowing us to report the validation and generation wall clock times separately. To fairly compare with \textsc{Leapfrog}, their sum can be considered the total wall clock running time of \textsc{Octopus}.
Across all benchmarks where \textsc{Leapfrog} terminates, \textsc{Octopus} outperforms it by several orders of magnitude in both runtime and memory usage.
In particular, the \emph{variable-length format~3} benchmark, which exceeded the machine's memory limits with \textsc{Leapfrog}, is successfully handled by \textsc{Octopus} within half an hour.
These results highlight the practical benefits of not constructing a proof object using elaborate L\textsubscript{tac} scripting.
Another interesting observation is that generation is often faster than validation.
We believe that this is the result of the successor checks being performed over larger formulas.

In the \emph{relational verification} and \emph{external filtering} benchmarks, the parsers are not strictly equivalent --- specifically, one accepts a strict subset of the packets accepted by the other.
The intention is that, for the more lenient parser, invalid packets are discarded at a later stage.
\textsc{Octopus}, like \textsc{Leapfrog}, allows refining the coherence check on Line~\ref{algo:coherence} of \cref{algo:bisim} to disregard such ``accepted'' packets that will be rejected later.
As a result, \textsc{Octopus} can establish behavioral equivalence of the two parsers modulo this predicate.

In terms of trusted base, \textsc{Octopus} does not produce a proof in the Calculus of Inductive Constructions, where \textsc{Leapfrog} does.
While our code is fairly simple, and a pen-and-paper proof of correctness for \cref{algo:bisim} is relatively straightforward, we have not formally verified the implementation.
In addition, correctness of \textsc{Octopus} relies on that of the underlying SMT solver, as well as correctness of the translation to SMT queries handled by \textsc{PySMT}.
While \textsc{Leapfrog} does formally verify the translation to a logic that can more-or-less be pretty-printed to an SMT query, the correspondence between the semantics of this logic and SMT queries is trusted.
Finally, \textsc{Leapfrog}'s trusted base also includes the underlying SMT solver, as the query response is not translated into a proof (\emph{à la} \textsc{SMTCoq}~\cite{ekici-etal-2017} or \textsc{CoqHammer}~\cite{czajka-kaliszyk-2018}) by their custom plugin.

\paragraph{Synthetic benchmarks.}
We now evaluate scalability using \textsc{Whippersnapper}~\cite{dang_whippersnapper_2017}, a synthetic parser generator.
We consider three classes of parsers.
\texttt{parse-field}, varying number of fields in a single header;
\texttt{parse-header}, varying the number of headers while keeping their size constant; and
\texttt{parse-complex}, varying parser depth and fan-out.
In all classes, a single header is parsed by a single state.
Each generated parser is compared against itself, where the expected outcome is equivalence --- although \textsc{Octopus} may decide negatively when different packets can be accepted depending on the (uninitialized) values in the store.
Experiments were run on the same server as the benchmarks in \cref{tab:experiment-findings}.
Runtimes are again averaged over three runs.

\begin{figure}[btp]
    \centering
    \includegraphics[
    width=1\linewidth,
    alt={%
    Three side-by-side plots showing Octopus's performance on synthetic benchmark sets: parse-field, parse-header, and parse-complex. The x-axes are the number of fields, headers, and states, respectively. The left y-axis shows runtime, and the right y-axis shows memory usage. Both runtime and memory increase steadily with input size, exhibiting approximately linear growth across all benchmarks.%
    }
    ]{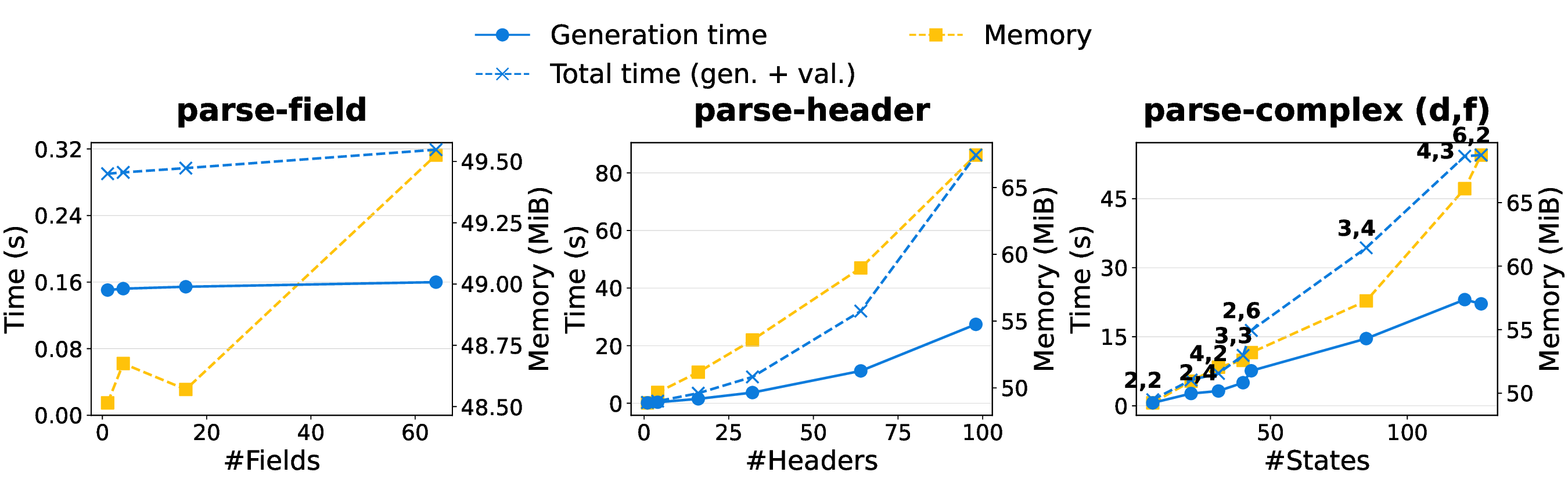}
    \caption{The total wall-clock time (left vertical axis, in seconds) and peak resident memory (right vertical axis, in MiB) required by \textsc{Octopus} running on synthetic benchmarks generated through \textsc{Whippersnapper}.}%
    \label{fig:whippersnapper}
\end{figure}

Results are shown in \cref{fig:whippersnapper}.
Increasing the number of fields has little impact on runtime, as it does not introduce new control-flow paths and can be handled efficiently using leaps.
By contrast, increasing the number of headers or the size of the parser tree substantially increases the number of execution paths, leading to corresponding growth in both runtime and memory consumption.
With an increasing number of execution paths, the respective formulas also grow in size.
Per the same reasoning as before, it thus makes sense that validation time grows faster than generation time.
For \texttt{parse-complex}, the number of states grows as \( \frac{f^{d+1}-1}{f-1} \), where $d$ is the depth and $f$ the fan-out, which is reflected in the observed scalability trends.
We consider these trends encouraging, as the largest parsers here already exceed the real-world parsers that we found publicly in terms of size and structural complexity.

We also verified several non-trivial equivalences between parsers drawn from the \texttt{parse-field}, \texttt{parse-header}, and \texttt{parse-complex} classes.
These checks succeeded within seconds, like earlier experiments.
For example, we showed an equivalence between a parser from \texttt{parse-header} with $n$ single-field headers and one with a single header of $n$ fields from \texttt{parse-field}, as well as between a \texttt{parse-complex} parser with a depth of $n$ and a fan-out of $1$, and a \texttt{parse-header} parser with $n$ headers.
This again demonstrates \textsc{Octopus}'s ability to reason about semantic, behavioral properties, rather than syntactic equivalence.

\paragraph{Public code.}
Finally, we investigate the performance of \textsc{Octopus} on code ``in the wild'', by examining a collection of publicly available P4 programs~\cite{DBLP:conf/sigcomm/HeBK019}.
This dataset consists of 34 programs from research projects, tutorials, and protocol implementations.
Among these, 30 fall within the scope of \textsc{Octopus} and can be read by \textsc{P4C}, potentially after porting their syntax to modern P4.
Of the remainder, two (\texttt{hyper4}, \texttt{axon}) contained header stacks, one more (\texttt{calc}) used lookahead, and another (\texttt{hashpipe}) was not accepted by our version of \texttt{P4C}.

We then compared each pair of parsers, leading to 435 equivalence checks.
We ran these experiments on a consumer laptop, with an Intel Core i5-12500H and $24$~GiB of memory. % chktex 8
On average, an equivalence check completed in $0.659$s (std: $0.261$s), with runtimes ranging from $0.300$ to $1.570$s and $50\%$ of checks completing between $0.430$ and $0.790$s.
Runtime measurements exhibited a mild right skew, with a few larger parser pairs accounting for the upper tail.
Peak resident memory usage averaged $51.72$~MiB (std: $4.34$~MiB), ranging from $46.40$ to $68.21$~MiB. Memory usage was more tightly concentrated around the median of $50.86$~MiB, with $50\%$ of runs falling between $49.96$ and $51.86$~MiB, again with a modest right tail corresponding to larger explored state spaces.

Our experiments identify two non-trivial equivalence classes: the first contains 4 parsers that do not decode further than an IPv4 header wrapped inside an Ethernet header; the second contains 6 parsers that go one step further, and may also decode a TCP header within this IPv4 header.
In both cases, the parsers in question are encoded in slightly different ways.
The remaining parsers are pairwise inequivalent, because they support different sets of protocols.

These measurements lead us to conclude that \textsc{Octopus} can handle parsers from various sources (as long as they are written in modern P4), and can furthermore successfully distinguish between heterogeneous parsers.

\section{Related Work}\label{sec:rel-work}
While \textsc{Octopus} is most closely related to \textsc{Leapfrog} (discussed in \cref{sec:introduction}), it fits into a broader body of work applying formal methods to P4 programs.

To support developers in verifying functional correctness, several push-button verifiers have been proposed; examples include \textsc{p4v}~\cite{liu_p4v_2018}, \textsc{Assert-P4}~\cite{neves_verification_2018}, and \textsc{Aquila}~\cite{tian_aquila_2021}.
These tools verify P4 programs against user-annotated functional properties by translating them into intermediate representations, which are then analyzed using formal verification techniques.
A complementary approach is taken by \textsc{Vera}~\cite{stoenescu_debugging_2018}, which can detect a wide range of practical bugs in P4 programs, without user annotations.
In contrast with all of these, \textsc{Octopus} does not reason about the (functional) correctness of a parser on its own.
Instead, it performs \emph{relational} verification by comparing the behaviors of two parsers.
Finally, a related line of work outside the P4 ecosystem is \textsc{EverParse}~\cite{DBLP:conf/uss/RamananandroDFS19}, which targets low-level binary message formats rather than programmable dataplane parsers.
Its goal is to synthesize parsers that are correct-by-construction, memory-safe, and non-malleable with respect to a given format specification.

A distinct class of tools targets the verification of translations to concrete network configurations, commonly referred to as \emph{tool stack verification}.
An example is \textsc{p4pktgen}~\cite{notzli_p4pktgen_2018}, which can generate a wide range of test cases to evaluate the correctness of a parser.
Such tools' goals are different from our own, which focus specifically on parser \emph{equivalence} rather than end-to-end correctness.

Most existing tools rely either on ad hoc semantic models or on existing implementations, such as \textsc{P4C}~\cite{noauthor_p4langp4c_2025}, the reference compiler for P4.
Efforts have also been made to provide formal semantics, such as \textsc{P4K}~\cite{kheradmand_p4k_2018}, which formalizes the $\text{P4}_{\text{14}}$ language specification in the K~framework~\cite{rosu_overview_2010}, and \textsc{Petr4}~\cite{doenges_petr4_2021}, which defines formal semantics for a subset of \pfour.

\section{Conclusion and Future Work}\label{sec:conclusion-future-work}
We presented \textsc{Octopus}, a tool that can verify equivalence of P4 parsers by symbolically checking bisimilarity of their automata, inspired by \textsc{Leapfrog}~\cite{doenges_leapfrog_2022}.
\textsc{Octopus} outperformed all test cases reported for \textsc{Leapfrog}, and managed an additional test case that was previously infeasible.
Using the \textsc{Whippersnapper} synthetic benchmark suite, we obtained a clear picture of \textsc{Octopus}'s scalability, and using the public code experiment, its applicability.
Together, these results indicate that \textsc{Octopus} is well-suited for real-world parser applications.

While \textsc{Octopus} exhibits practical runtime performance, there remains potential for further improvement.
As observed in~\cref{sec:eval}, certificate validation approximately doubles execution time.
One direction for future work is to embed certificates for generated SMT queries and validate those, instead of re-running the SMT solver.
This could substantially reduce overall execution time.

We are also interested in extending the scope of \textsc{Octopus}.
Some extensions appear to require straightforward engineering effort, while others, such as supporting lookahead together with termination detection, likely require additional theoretical development.
We are currently exploring a preliminary implementation of the latter.
Such extensions could enable support for a broader range of real-world parsers, including the P4 fabric used by ONOS~\cite{onos}.

\begin{credits}
\subsubsection{\ackname}\label{sec:acknow}
This work is based on the first author's bachelor’s thesis~\cite{van_leenen_practical_2025}.
The authors would like to thank Jan Martens for his comments on that version, and Emily Yu for her helpful comments on a later draft of this manuscript. We also wish to thank the anonymous reviewers for their helpful comments and suggestions.

This work was partially supported by the Dutch Research Council (NWO) under grant no. VI.Veni.232.286 (ChEOpS).

\subsubsection{\discintname}
The authors have no competing interests to declare that are relevant to the content of this article.

\end{credits}

\section*{Data-Availability Statement}
The data and artifact supporting the results of this study are publicly available via Zenodo at \url{https://doi.org/10.5281/zenodo.19832336}. The artifact is provided as a Docker image and enables reproduction of all reported results, except for the runtime measurements of \textsc{Leapfrog} in~\cref{tab:experiment-findings}, which are excluded due to their computational cost. The artifact and the tool's source code are released under the MIT License.

\bibliographystyle{splncs04}
\bibliography{octopus}

\end{document}